\documentclass[prc,preprint]{revtex4}
\usepackage{graphicx}
\usepackage{subfigure}
\usepackage{amsmath,amssymb}
\usepackage{mathrsfs}
\def\dis{distribution}
\def\fq{$F_q$}
\def\pt{p_T}

\def\ps{\psi_q(p)}
\def\cp{C_{p,q}}
\def\cpq{$C_{p,q}(M)$}

\def\mq{\mu_q}

\def\sg{\Sigma_q}
\def\pr{\propto}
\def\bq{\begin{eqnarray}}
\def\eq{\end{eqnarray}}
\begin{document}
\title{Observable Properties of Quark-Hadron Phase Transition at the Large Hadron Collider}
\author
 {Rudolph C. Hwa$^1$ and C.\ B.\ Yang$^{2}$}
\affiliation
{$^1$Institute of Theoretical Science and Department of
Physics\\ University of Oregon, Eugene, OR 97403-5203, USA\\
\bigskip
$^2$Key Laboratory of Quark and Lepton Physics\\  Central China Normal University,
Ministry of Education, P.\ R.\ China}
\date{\today}
\begin{abstract}
Quark-hadron phase transition is simulated by an event generator that incorporates the dynamical properties of contraction due to QCD confinement forces and randomization due to the thermal behavior of a large quark system on the edge of hadronization. Fluctuations of emitted pions in the $(\eta,\phi)$ space are analyzed using normalized factorial moments in a wide range of bin sizes. The scaling index $\nu$ is found to be very close to the predicted value in the Ginzburg-Landau formalism. The erraticity indices $\mu_q$ are determined in a number of ways that lead to the same consistent values. They are compared to the values from the Ising model, showing significant difference in a transparent plot.  Experimental determination of $\nu$ and $\mu_q$ at the LHC are now needed to check the reality of the theoretical study and to provide guidance for improving the model description of quark-hadron phase transition.

\end{abstract}
\maketitle

\section{Introduction}
Critical phenomenon is a subject of interest in many areas of physics because of its universality. It has been exhaustively studied for condensed matters at low temperature, but hardly considered at high temperature, such as for cosmic phase transition in the early universe, due to the scarcity of data. The Large Hadron Collider (LHC) provides a unique opportunity to study the properties of quark-hadron phase transition at high temperature because the Pb-Pb collision energy is high enough to produce not only quark-gluon plasma (QGP) on the one hand, but also thousands of hadrons on the other to allow severe cuts in data analysis to isolate fluctuation patterns that characterize critical phenomena. It is well known that the tension between collective interaction and thermal randomization near the critical temperature results in clusters of all sizes. The observation of such patterns that are scale-independent is therefore the primary objective of a search for revealing signatures of critical behavior.

Specific measures for detecting the scaling properties in heavy-ion collisions were proposed many years ago \cite{hn, cgh, rh}. They involve the use of the Ginzburg-Landau (GL) theory of second-order phase transition (PT) \cite{gl} and of the Ising model \cite{kh} to simulate spatial patterns. The results obtained in such theoretical studies will be useful in interpreting experimental results from analyzing heavy-ion collision data. However, the GL description is a mean-field theory and the Ising model provides only static geometrical patterns. They lack the dynamical content of how quarks turn into hadrons, as the QGP becomes dilute enough for the confinement force to come into play. Having just mentioned hadronization of quarks, it is immediately necessary to caution the inadequacy of the usual mechanisms (such as Cooper-Frye \cite{cf}, parton fragmentation or recombination) for treating the problem of PT, because what we need is a description of the global properties of a two-dimensional surface through which the quarks inside emerge at some places as hadrons at certain times, but remain unconfined at other places until a later time when confinement occurs. If such fluctuations are the consequences of the dynamics of PT, then on one hand we expect by virtue of the universality of critical phenomena that the results from GL and Ising considerations are still relevant, while on the other hand it is desirable to have a model that incorporates the confinement dynamics in the final stage of the evolution of the quark system. It is the aim of this paper to discuss both of these two aspects of the problem so that what can be measured at the LHC can shed light on the traditional treatment of PT, and vice versa.

The study of scaling behavior of fluctuations in geometrical configurations in multiparticle production is usually done by use of the factorial moments \cite{bp} and recognized in terms of a phenomenon referred to as intermittency \cite{bp1}. Such studies have been carried out earlier by experiments at CERN, Brookhaven and JINR \cite{ddk, kw}, but no definitive conclusions could be drawn on critical behavior because collision energies were not high enough to facilitate selection of relevant events that could clearly exhibit crucial characteristics without being smeared out by the averaging procedures necessitated by inadequate statistics. At LHC the particle multiplicities are so high that narrow $\pt$ intervals can be chosen while still allowing bin size in the $(\eta,\phi)$ space to vary over a large range so as to reveal scaling properties. When that is done experimentally, the burden of further progress shifts to the theoretical side in search for pertinent measures that go beyond intermittency. Erraticity is such a measure constructed from moments of the factorial moments \cite{ ch, ch1,rh1,ch2,ch3}. An analogy of the relationship between intermittency and erraticity is that between the mean and width of a peaked probability distribution. The model that we shall describe is capable of probing the detailed properties of erraticity, thereby offering a view of critical behavior that has never been considered before in the traditional treatment of such phenomena.

\section{Ginzburg-Landau, Ising and Quark-hadron Phase Transition}

We begin with a summary of what has been done on the subject so that we have a clear framework to start out from.
In heavy-ion collisions we consider the surface of a cylinder that contains the QGP at a particular time in the late stage of its evolution. That cylinder is in the $(\eta,\phi)$ space of the emitted hadrons; the remaining variable $\pt$, the transverse momentum, is orthogonal to $(\eta,\phi)$ and is the only observable that can in some approximate, though certainly not unique, way be related to the time of hadronization. 
That is based on the premise that the emission of jets with high $\pt$ occurs early, soon after collision, and that the thermalized plasma with high pressure gradient emits intermediate-$\pt$ hadrons earlier than the more dilute plasma emitting the low-$\pt$ hadrons near the end of its hydrodynamical expansion.
We shall make cuts in $\pt$  at low $\pt$ with small $\Delta\pt$ interval to examine the fluctuation patterns in a square lattice, which can be mapped to the $(\eta,\phi)$ space through the use of cumulative variable \cite{bg}. The unit square is divided into $M_1\times M_2$ bins, and we shall use $M$ to denote the total number of bins, $M_1M_2$. 

The normalized factorial moment $F_q$ is defined by 
\bq
F_q=\left< {f_q\over f_1^q}\right>_v,  \qquad \qquad  f_q=\sum_{n=q}^\infty {n!\over (n-q)!} P_n ,  \label{1}
\eq
where $n$ is the number of particles in a bin and $P_n$ is its distribution over all $M$ bins. Thus 
$f_q=\left< n(n-1)\cdots (n-q+1)\right>_h$ is the average over all bins of a given event, where the subscript $h$ implies {\it horizontal} in the sense that different events are stacked up vertically. \fq\ is then the {\it vertical} average over all events. As has been pointed out in Ref.\ \cite{bp, bp1} and later reviewed in \cite{rh, kw}, \fq\  has the important property that statistical fluctuations in $P_n$ are filtered out so that only non-trivial dynamical fluctuations can result in \fq\ being different from unity. Intermittency refers to the scaling behavior
\bq
F_q \pr M^{\varphi_q}.  \label{2}
\eq
That behavior has been observed in many experiments \cite{na, em, emu, es, kw}.

To examine the effects of quark-hadron PT on the observable patterns in the $(\eta,\phi)$ space, it is illuminating to apply the Ginzburg-Landau formalism \cite{gl} to the calculation of \fq. Without giving the details that can be found in \cite{hn, rh}, we state here the result; i.e., \fq\ has the scaling behavior, referred to as $F$-scaling,
\bq
F_q \pr F_2^{\beta_q} ,  \label{3}
\eq
for a wide range of $M$. The exponent $\beta_q$ satisfies
\bq
\beta_q=(q-1)^{\nu},  \qquad \nu=1.304 .   \label{4}
\eq
Although the GL parameters are dependent on the temperature $T$, 
the index $\nu$ is independent of the details of the GL parameters so long as  $T$ is less than the critical temperature $T_c$. To have a numerical value for $\nu$ is highly desirable, especially in view of the fact that $T$ is not a variable under experimental control in heavy-ion collisions. The beauty of the uniqueness of $\nu$ in the GL theory is also its drawback in that it does not inform us about the nature of the physical system if a measurement of $\nu$ yields a value that is in the proximity of 1.3, but not exactly at the value in Eq.\ (\ref{4}). Being independent of $T$, it means that $\nu$ is a value averaged over all $T$ at which the PT can take place. This point will be made visually non-trivial in Fig.\ 4 below.

The spatial pattern of where hadrons are produced in the $(\eta,\phi)$ space can best be simulated in the 2D Ising model, which is known to exhibit second-order PT \cite{kh, bdf}. The model has near-neighbor interaction that generates collective ordered behavior and thermal motion that generates random disordered behavior.
It is known that the Ising model leads to intermittency behavior at the critical point \cite{hs}. 
 The application to hadron production in heavy-ion collisions has been carried out in Ref.\ \cite{cgh}, in which the net spin up in a small cell in the Ising lattice is identified with the presence of hadrons and net spin down with no hadrons, where spin up or down is defined with reference to the overall magnetization of the whole lattice. More precisely, the hadron multiplicity is proportional to the absolute square of the local mean magnetic field of the cell; the proportionality constant $\lambda$ is a scale factor that relates the quark density of the plasma at $T_c$ to the lattice site density in the Ising model. Since $\lambda$ relates a physical space to a mathematical space, any observable consequences implied by the model calculation should be insensitive to the value of $\lambda$. In the model one can vary $T$ so that at $T<T_c$ more spins are aligned due to the dominance of the collective force that is ordered, while at $T>T_c$ the lattice spins are more likely to be misaligned due to the disordered nature of thermal randomization. It is found in \cite{cgh} that the value of $T_c$ can be well determined by studying the $M$-scaling behavior in Eq.\ (\ref{2}), which occurs over the widest ranges of $M$ and $\lambda$ at $T=T_c$. In terms of $J/k_B$, where $J$ is the strength of interaction between nearest neighbors of spins on the Ising lattice and $k_B$ the Boltzmann constant, $T_c$ is found to be 2.315, only slightly higher than the analytical value of 2.27 for infinite lattice \cite{kh}.

After establishing a connection between the Ising model and hadron counting through the use of \fq, it is then meaningful to examine $F$-scaling for the Ising configurations for a range of $T<T_c$. It is found in \cite{cgh} that the scaling exponent $\beta_q$ satisfies the power-law behavior in Eq.\ (\ref{4}), but with the index $\nu$ being dependent on $T$. That dependence, shown in Fig.\ 11 of \cite{cgh}, provides an interpretation of the observable quantity $\nu$ in heavy-ion collisions in terms of an aspect of quark-hadron PT that depends on $T$. As we shall exhibit in Fig.\ 4 below, the GL value of $\nu=1.304$ is an average of the Ising values $\nu(T)$ between $\nu=1.04$ at $T=T_c=2.315$ and $\nu=1.56$ at $T=2.2$.

	To summarize what we have reviewed above, the GL description of second-order PT is a mean-field theory, in the framework of which the $F$-scaling properties are independent of $T$. The Ising model exhibits explicitly spatial configurations at any $T$; thus the scaling index $\nu$ calculated from those configurations depends on $T$. The variation of $\nu(T)$ for $T<T_c$ is consistent with the GL value and provides an insight into the temperature of a system that undergoes a PT with a specific value of $\nu$.

	To proceed further it is clear that we need a model that contains in some way the specific nature of QCD dynamics that can give rise to quark-hadron PT when the QGP is dilute enough for hadrons to form. Because of the complexity of the system both at the global level where cooperative phenomenon occurs and at the local level where specific process of hadronization takes place, in addition to the problems of time evolution of average density on one hand and fluctuations from the average on the other, drastic approximations shall be made to construct an event generator that can capture the essence of the characteristics responsible for PT of the quark system. We shall describe such an event generator in the next section. It will be referred to as SCR. Using it to generate configuration in a square lattice, as has been done in the Ising model, we can subdivide the unit square into $M=M_1M_2$ bins as described in the first paragraph of this section, and then calculate \fq\ in accordance to Eq.\ (\ref{1}). Without digressing to describe the details of the event generator, we show in Fig.\ 1 the dependence of \fq\ on $M$ for $q=2, \cdots, 5$, and for $\pt=0.2$ GeV/c with an interval $\Delta\pt=0.1$ GeV/c around it. We note that there are two scaling regions separated by a transition region between $M=5\times 10^2$ and $4\times 10^3$. 
\begin{figure}[tbph]
\vspace*{0.5cm}
\includegraphics[width=.7\textwidth]{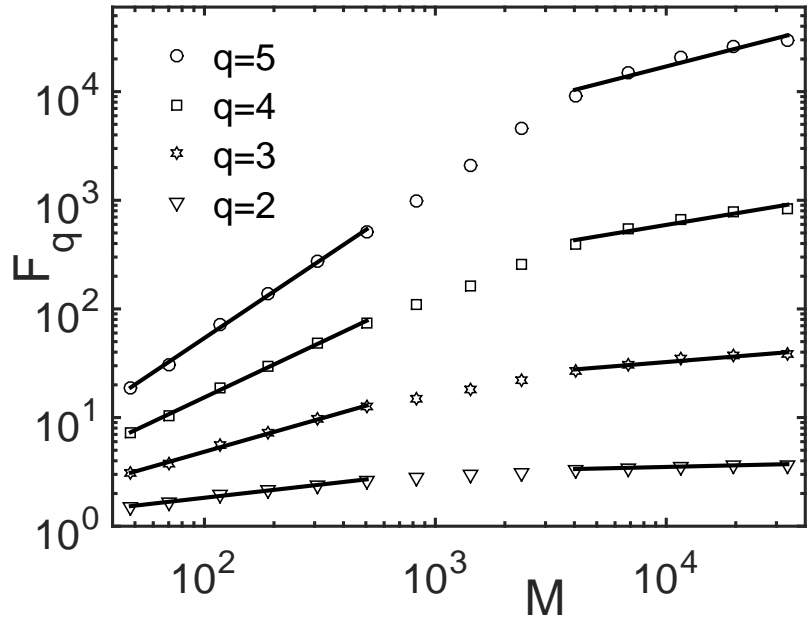}
\caption{Scaling behavior of $F_q(M)$}
\end{figure}
\begin{figure}[tbph]
\vspace*{0.5cm}
\includegraphics[width=.7\textwidth]{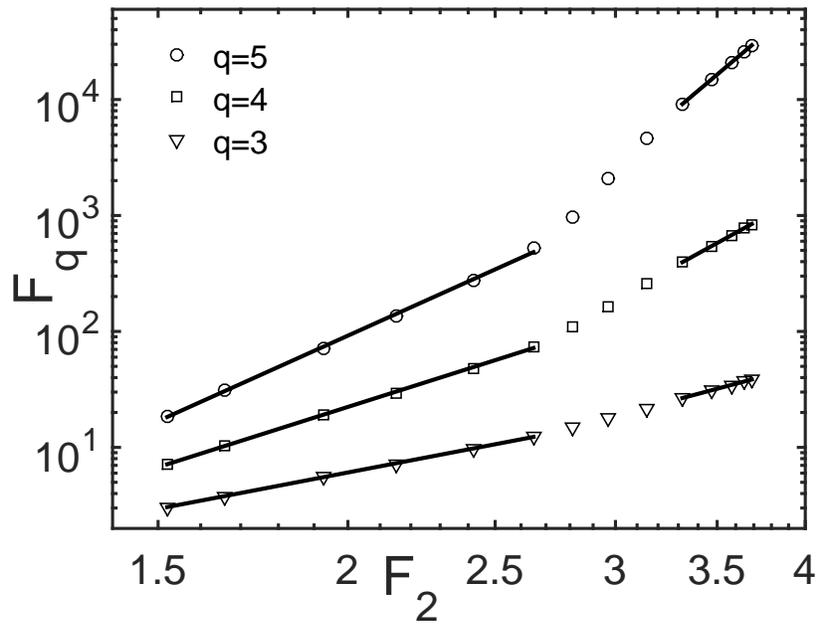}
\caption{$F$-scaling behavior of $F_q$ vs $F_2$.}
\end{figure}	
\begin{figure}[tbph]
\vspace*{0.5cm}
\includegraphics[width=.7\textwidth]{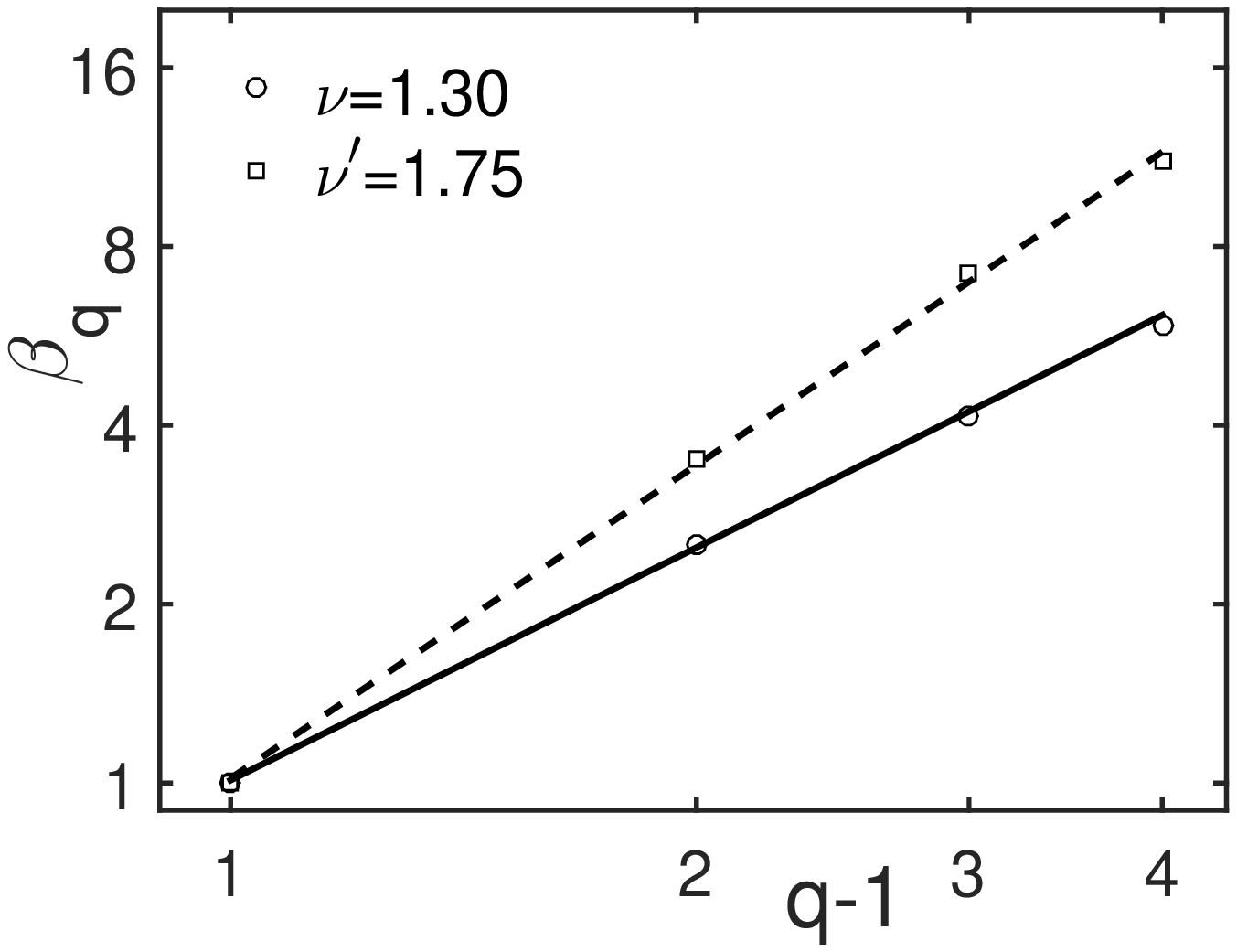}
\caption{Log-log plot of $\beta_q$ vs $q-1$. $\nu$ and $\nu'$ are the values of the slopes.}
\end{figure}	
	The corresponding plot of $F$-scaling is shown in Fig.\ 2, in which the upper region is very short. The major scaling region is on the low side, exhibiting sustained linear behavior in the log-log plot. From their slopes we determine the exponents $\beta_q$ in Eq.\ (\ref{3}). The power-law dependence of $\beta_q$ on $q-1$ is shown by the solid line in Fig.\ 3 with the index $\nu$ determined to be
\bq
\nu=1.30 .  \label{5}
\eq
The upper scaling region, though short, nevertheless has a power-law behavior also, as shown by the dashed line in Fig.\ 3. The corresponding value of the $\nu$-index is
\bq
\nu'=1.75.   \label{6}
\eq
These are the first results on $\nu$ based on a model treatment of quark-hadron PT due to confinement forces.

The lower scaling region (for which $M<500$) exhibits the properties of spatial fluctuations when examined in coarse-grain analysis without going into the details of sharp spikes at high resolution. Those fluctuations correspond very well to the 
 fluctuating patterns of the Ising configurations in which the net spin in the direction of the overall magnetization in a small cell consisting of several lattice sites is identified with a presence of a hadron. The scale-independent spatial patterns in the extended coordinate space do not require a large number of bins to isolate narrow peaks in small bins. Thus the lower scaling region is a true measure of the quark-hadron PT that corresponds to the Ising fluctuations, and the value $\nu=1.3$ agrees excellently with the GL result.

	The higher scaling region suggests the presence of sharp spikes in small bins that would not show up in a mean-field theory such as GL. It turns out to be rich in physics content and will be investigated in detail in Sec.\ IV below. For the present purpose, we postpone all discussions on that subject until later.

	We remark that different scaling behaviors were observed in different regions two decades ago \cite{es}, which were interpreted as possible evidence for non-thermal phase transition \cite{brp, ggs, ddk, kw} that is different from the usual one \cite{rp, bz, brp2}. Due to the drastically different collision energies, multiplicities, bin numbers, dimensions and methods of analyses, it is not clear whether there is any parallelism in the phenomena found there and here, apart from the recognition that both have multiple scaling behaviors. 

\begin{figure}[tbph]
\vspace*{0.5cm}
\includegraphics[width=.7\textwidth]{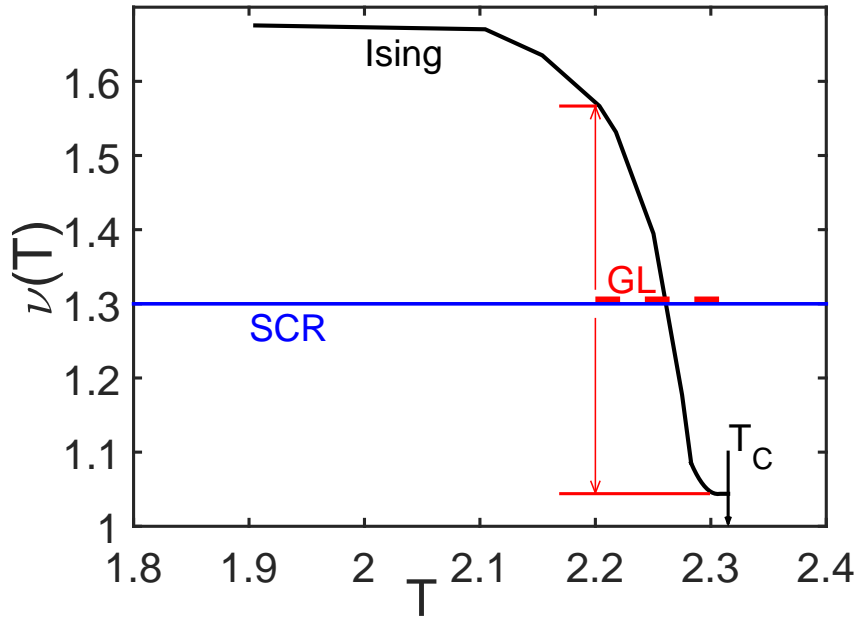}
\caption{(Color online) Values of $\nu(T)$ from three models: (a) Ising (black), (b) Ginzburg-Landau (red), (c) SCR (blue). The critical temperature in the Ising model is $T_c=2.315$ in Ising units. The GL value in dashed red line may be regarded as the average of the Ising values between $T=2.2$ and $T_c$.}
\end{figure}	

	It is now opportune to put together the results from GL, Ising and SCR by showing in Fig.\ 4 the various values of $\nu$. The solid (black) line is adapted from Fig.\ 11 of Ref.\ \cite{cgh}; it is calculated in the framework of the Ising model where $T$ is a control parameter. The scale factor $\lambda$ that relates the quark density to the Ising lattice site density is not shown in Fig. 4 here for clarity's sake, since the result is essentially independent of $\lambda$. The curve $\nu(T)$ provides a model-dependent interpretation of the value of $\nu$ in terms of temperature in the sense that: (a) critical behavior can occur only at $T<T_c= 2.315$ in Ising units; (b) at $T$ progressively less than $T_c$ the dynamical fluctuations measured by \fq\ become more dominant, resulting in larger $\nu$, and (c) the dominance of collective behavior cannot continue at ever-lower $T$ because critical phenomena depend on the balanced tension between the collective and the random forces. The red dashed line is the GL value in Eq.\ (\ref{4}) and is an average between 1.04 and 1.56, corresponding to $2.2<T<2.315$. As has been described in Ref.\ \cite{hn, rh}, the $\nu$ value in the GL description of PT is insensitive to the GL parameters (and thus to the temperature) so long as $T<T_c$.

	Our value of $\nu$ in Eq.\ (\ref{5}) from SCR is represented by the horizontal (blue) line at 1.3 in Fig.\ 4. Although SCR keeps track of the temperature by bookkeeping in order to generate a $\pt$ \dis, the randomizing procedure in event generator does not depend on the plasma surface temperature at each time step. Thus, we cannot meaningfully assign any $T$ dependence or $T$ interval to our result on $\nu$ from SCR. It is, nevertheless, amazing that by incorporating contraction and randomization subprocesses into SCR the resultant $\nu=1.3$ can come so close to the GL value of $\nu=1.304$ that is not associated with any specific dynamics.
	
	It should be noted that there have been experiments at lower energies where $\nu$ is found to be higher than 1.3 \cite{kw, vas, aa}. Those results are based on 1D analysis of fluctuations in the $\eta$ space only without cuts in either $\phi$ or $\pt$. The maximum numbers of bins are not very high compared to what we consider. It is unclear what features of the fluctuations have been smeared out by the averaging process in $\phi$ and $\pt$. Thus their results cannot be put on Fig.\ 4 to infer any conclusion about whether a QGP has been formed or a second-order PT has occurred. 
	As matters stand at this point on the subject of  scaling index $\nu$ for PT, Fig.\ 4 provides a satisfactory summary of what  are known theoretically, and needs only an experimental input from LHC to shed light on their reality.

\section{Event Generator SCR}

	We now describe the event generator SCR, which stands for Successive Contraction and Randomization. The general idea behind it is  described in Ref.\ \cite{hy}, but because of some changes in parameters we outline here the step-by-step procedure of the simulation algorithm, which corresponds to the critical case in \cite{hy}.

\noindent {\bf 1.	 Initial Configuration}

A unit square $S$ is seeded initially with 1,000 $q\bar q$ pairs in the form of clusters with probability distribution $P(C)\pr C^{-2}$ such that $C$ pairs of $q\bar q$ are grouped together in a cluster centered at a random point in $S$. The distribution within the cluster, with $q$ and $\bar q$ being independent, is Gaussian around the center with a width
\bq
\sigma=0.1 C^{-1/2} .  \label{7}
\eq
We stop the seeding process when the total $q\bar q$ pairs reaches 1000. Since the unit square $S$ is mapped onto the $(\eta,\phi)$ space at mid-rapidity, we assign an initial temperature to the initial configuration at $T = 0.4$ GeV, so that the $q$ and $\bar q$ can independently be given a value of $\pt$ in accordance with the thermal distribution $\exp(-\pt/T)$. This is the zeroth step at $t_0=0$.

\noindent {\bf 2.	Pionization}

If a $q$ and $\bar q$ are separated by a distance $d$ that is less than $d_0=0.03$, we regard the confining force to be effective in recombining the pair to form a pion regardless of color and flavor, whose attributes we ignore in this study. We record the position in the square (midpoint of the pair) and the momentum of the pion (sum of the $\pt$ of $q$ and $\bar q$), and remove the $q\bar q$ pair from $S$. A pion is thereby emitted from the cylindrical surface. This is done for all pairs that are close enough for recombination.

\noindent {\bf 3.	 Contraction}

To implement the global effect of confinement forces rather than just between nearby color charges, we devise the contraction scheme that is central to the collective behavior of an extended system. We divided $S$ into $5\times 5$ bins and separate the 25 bins into two groups: dense and dilute. The dense bins have more multiplicity per bin (counting $q$ and $\bar q$ independently) than the average. If neighboring dense bins share a common side, they are grouped together to form a cluster of dense bins. Let $D$ refer to such a cluster of connected dense bins, and $N_D$ the number of bins in $D$. Define $\vec r_D$ to be the coordinates in $S$ that is the center of mass of $D$. A contraction of $D$ is a redistribution of all $q$ and $\bar q$ in D, centered at $\vec r_D$, but with a Gaussian width
\bq
\sigma_D=0.05 N_D^{1/2} .   \label{8}
\eq
Since the redistribution puts the $q$ and $\bar q$ mostly under the Gaussian peak, the process represents a contraction of the particular cluster $D$. There are, however, many other clusters in $S$. The same procedure leads to contraction of each and every one of them to their respective centers. That means there will be more dilute bins before the next dynamical action. But first we allow pionization to take place in the contracted configuration. 

\noindent {\bf 4.	 Randomization}

Thermal randomization is the disordered motion that opposes the ordered collective motion. We implement that aspect of the opposing force by requiring all $q$ and $\bar q$ in the dilute bins to be redistributed randomly throughout $S$. Note that this step is not coordinated with the temperature of the system, which is a procedure that seems more complicated than is worthwhile at this stage. After randomization we return to the previous step of contraction and pionization. We continue this iterative process of contraction, pionization and randomization until 95\% of the $q\bar q$ system is depleted. That is regarded as the end of one time step in which nearly all the quarks on the plasma surface have undergone a quark-hadron PT. The next layer of quarks in the plasma interior then moves out to the surface, so we proceed to the next time step.

\noindent {\bf 5.	Subsequent Time Steps}

	At each time step $t_i, i=1,2,\cdots$, we add 200 new pairs of $q\bar q$ to $S$ that contains the remnants from the previous step. The new pairs are distributed according to $P(C)$ as before, and have $\pt$ distribution with an inverse slope
\bq
T_i=T_{i-1}-0.02\ {\rm GeV}.  \label{9}
\eq
Thus the temperature is lower at later time in keeping with the general notion of hydrodynamical expansion. We then recycle the steps of contraction, pionization, and randomization repeatedly until 5\% of the  $q$ and $\bar q$  remains before moving on to the next time step. This process continues for 10 time steps, so the total number of $q\bar q$ pairs introduced to the system is 3000. The final $\pt$ distribution for $\pt<1$ GeV/c is approximately exponential with an inverse slope of 0.285 GeV/c. Our aim is not to fit the experimental $\pt$ \dis, but to obtain configurations in the $(\eta,\phi)$ space for any sensible cut in $\pt$. That is what we have accomplished in SCR.

\section{Erratic Fluctuations}

	We have seen in Sec.\ II that the factorial moments \fq\ can effectively describe the fluctuations of spatial configurations through their scaling behaviors, when the system undergoes a second-order PT, whether the system is a 2D Ising lattice or a simulated quark system near hadronization. In the case of the Ising model it is clear that each configuration involves clusters of various sizes with spins pointing up or down (and we identify only spin-up relative to the overall net magnetization with non-vanishing hadron density). While all those configurations are different from one another, they are all rather similar in their main characteristics. It means that  the probability distribution, $P_n$, of $n$ particles in a bin introduced  in Eq.\ (\ref{1}) may   be narrow for the range of $M$ studied. Alternatively, it may just be that  the $f_q$ for each event (i.e., configuration) is such that $f_q/ f_1^q$ in (\ref{1}) does not fluctuate too much from event to event.

	To learn more about multiplicity fluctuations, we need a measure that is sensitive to the width of $P_n$. To that end we consider the moments of moments. For an event $e$, the horizontal factorial moments are 
\bq
F_q^e(M)=f_q^e(M)/[f_1^e(M)]^q ,   \label{10}
\eq
whose vertical average is just \fq, as shown in Eq.\ (\ref{1}). If, at large $M$, $F_q^e(M)$ does not vanish even at $q = 5$, say, that must mean the existence of a spike in some bin where $n\ge q$ even though the average $\left< n\right>$ per bin may be miniscule. Such a value of $F_q^e(M)$ must deviate strongly from the vertical average $\left<F_q(M)\right>_v$, and represents the type of erratic fluctuations that we want to quantify. We note that such erraticity cannot happen in the analysis of the Ising model because of the definition of hadrons in terms of cells that are not extremely small \cite{cgh}. To focus on the deviation of $F_q^e(M)$ from $\left<F_q(M)\right>_v$, let us define 
\bq
\Phi_q(M)=F_q^e(M)/\left<F_q(M)\right>_v ,   \label{11}
\eq
and consider the $p$th power of $\Phi_q(M)$ before averaging, i.e.,
\bq
\cp(M)=\left<\Phi_q^p(M)\right>_v ,   \label{12}
\eq
where $p$ is $\ge 1$, but need not be an integer. Clearly, with large $p$ the events with large $F_q^e(M)$ make more important contribution to the vertical average, and that probes more into the large $n$ tail of $P_n$ when $M$ is large. If $\cp(M)$ has a scaling region in which it behaves as
\bq
\cp(M) \pr M^{\psi_q(p)} ,  \label{13}
\eq
the phenomenon is referred to as erraticity \cite{rh1, ch, ch1, hy}. Compared to Eq.\ (\ref{2}), it evidently represents a step beyond intermittency. Since at $p = 1, C_{1,q}(M)=1$, so $\psi_q(1)=0$, any non-vanishing  $\psi_q(p)$ is a window toward a new territory in fluctuations.

	In the situation where $\ps$ depends linearly on $p$ (a case which we shall show to be generated by SCR), then the slope at $p = 1$ carries  information beyond $p=1$. We define 
\bq
\mu_q={d\over dp}\ps|_{p=1}   \label{14}
\eq
and refer to it as an erraticity index that is independent of $M$ and $p$. We advocate the use of $\mu_q$ as a measure of the dynamical fluctuations in heavy-ion collisions. On the one hand, it is observable at the LHC, while on the other hand it can be related not only to quark-hadron PT \cite{hy} but also to classical chaos \cite{ch2,ch3}.

\begin{figure}[tbph]
\vspace*{0.5cm}
\includegraphics[width=.7\textwidth]{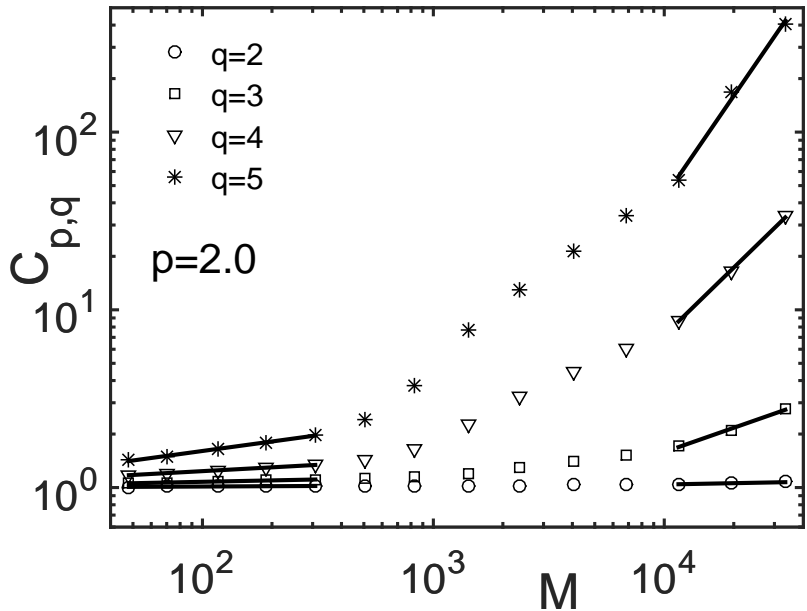}
\caption{Scaling behaviors for $C_{p,q}(M)$.}
\end{figure}

	To study the $M$ dependence of $C_{p,q}(M)$ for $p>1$, we see in Fig.\ 5 that there are two scaling regions: (a) $M<300$, and (b) $M>10^4$, where straight lines are drawn connecting the points in those regions. It is also possible to identify a linear region in between those two regions, but no attention will be given to that region in this paper. Figure 5 is for $p=2$, and is representative of the behavior of $C_{p,q}(M)$ at other values of $p$. The two scaling regions roughly correspond to the lower and upper regions seen in Fig.\ 1. In Sec.\ II we have investigated the intermittency behavior in the lower region. We now give attention to the upper region where sharp spikes in small bins can give rise to erraticity. But first we focus on the immediate neighborhood of $p=1$. We cannot present a figure like Fig.\ 5 for $p=1$ because $C_{1,q}(M)=1$  identically. The closest to it would be the derivative of $C_{p,q}(M)$ at $p=1$. From Eq.\ (\ref{12}) we have
\bq
	\sg(M)={d\over dp}\cp(M)|_{p=1} = \left<\Phi_q \ln \Phi_q\right>_v  \label{15}
\eq
where the notation of using $\sg$ suggests entropy due to the last expression above. Although the connection with entropy \cite{ch2, ch3} is not of crucial importance here, its connection with $\mq$ is directly relevant since the scaling behavior in Eq.\ (\ref{13}) implies
\bq
\sg(M)\pr {d\over dp}M^{\psi_q(p)}|_{p=1} = \mq \ln M .  \label{16}
\eq
The above expression is not to be taken to mean that $\Sigma_q(M)$ is proportional to $\ln M$ in general. Equation (\ref{13}) is an expression of the scaling behavior (for $M$ in the upper scaling region in our present  consideration) without the implication that there can be no constant background, so also in Eq.\ (\ref{16}) $\Sigma_q(M)$ may have the form of $\sigma_q+\mu_q\ln M$ at large $M$. In Fig.\ 6 we show the result on $\sg$ vs $\ln M$ from SCR. We can identify the last three points at $M>10^4$ as showing some degree of linear behavior. However, when we plot $\sg$ vs $\Sigma_2$ as in Fig.\ 7, the linear region becomes quite extensive.
Denoting the slope in general by $\omega_{q_0}(q)$, we have
\bq
\omega_{q_0}(q)={\partial \sg\over \partial \Sigma_{q_0}}=
{\partial \Sigma_q(M)/\partial \ln M\over \partial \Sigma_{q_0}(M)/\partial \ln M}={\mu_q\over \mu_{q_0}} .   \label{17}
\eq
It should be noted that SCR has not been tuned to fit any real data from LHC, but it does provide concrete representations of what we have discussed so far in theoretical terms.

\begin{figure}[tbph]
\vspace*{0.5cm}
\includegraphics[width=.7\textwidth]{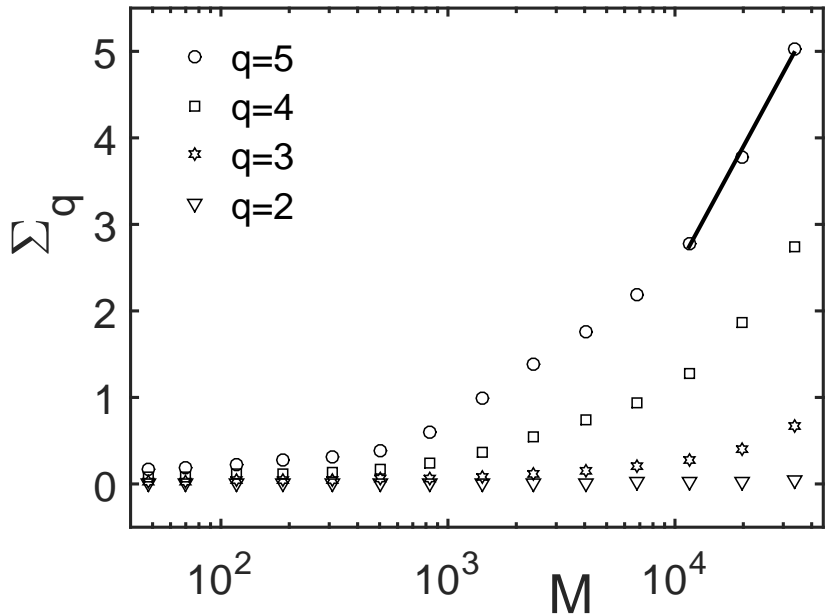}
\caption{Semilog plot of the $M$ dependence of the entropy function $\Sigma_q(M)$.}
\end{figure}	
\begin{figure}[tbph]
\vspace*{0.5cm}
\includegraphics[width=.7\textwidth]{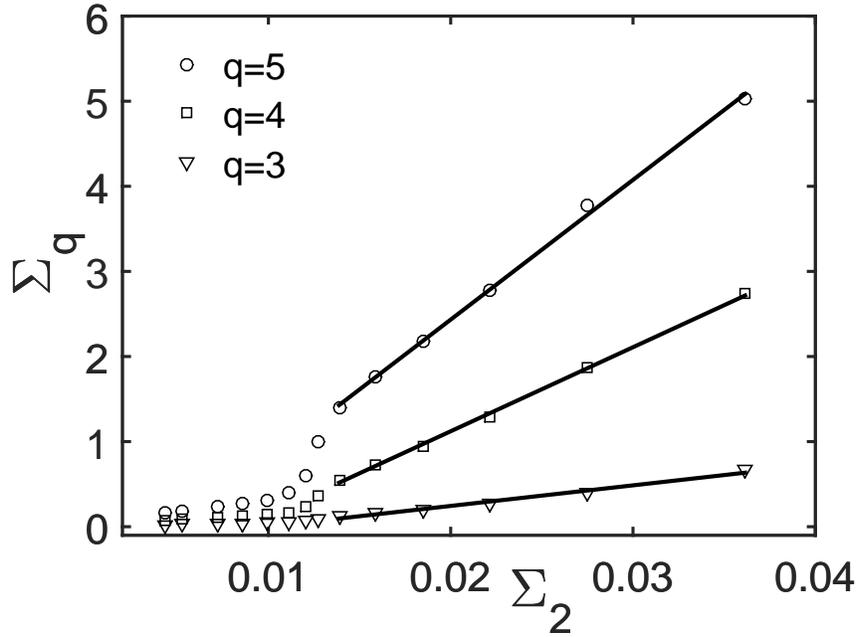}
\caption{Linear plot of $\Sigma_q$ vs $\Sigma_2$.}
\end{figure}	

	
	The  slopes in Fig.\ 7 are (for $q_0=2$)
\bq
\omega_2(q)=1.0, 24.2, 98.8, 164.2   \qquad {\rm for}  \qquad q=2,3,4,5.   \label{18}
\eq
To determine $\mq$ we rely on Eq.\ (\ref{16}) and Fig.\ 5 and obtain for $q=5$
\bq
 \mu_5= {\partial \Sigma_5(M)\over \partial \ln M} = 2.1 .    \label {19}
 \eq
From Eq.\ (\ref{17}) we get with the help of (\ref{18})
\bq
	 \mu_2=\mu_5/ \omega_2(5)=0.0128 ,  \qquad 
	 \mu_3=\mu_2\omega_2(3)=0.309, \qquad \mu_4=\mu_2\omega_2(4)=1.265 .  \label{19c} 
\eq

\begin{figure}[tbph]
\vspace*{0.5cm}
\includegraphics[width=.7\textwidth]{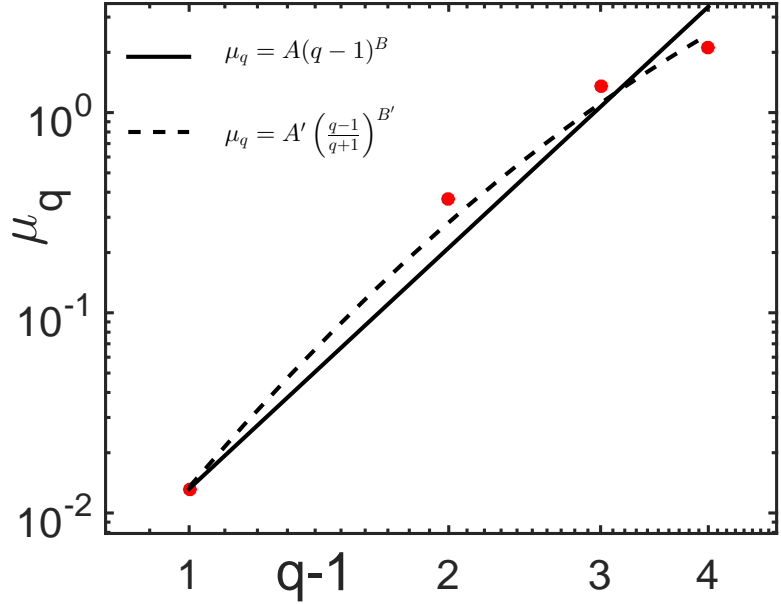}
\caption{(Color online) $q-1$ dependence of the erraticity index $\mu_q$. The red points are from Eqs.\ (19) and (20). The solid and dashed lines are fits by the equation shown.}
\end{figure}

\noindent These values of $\mq$ are shown in Fig.\ 8. One may choose a formula that can fit all four points in that figure. However, for a reason that will become self-evident shortly, let us choose a simple formula
\bq
\mu_q=A(q-1)^B  \label{20}
\eq
that can approximate the points of $\mq$ by a straight line in Fig.\ 8 with just two parameters $A$ and $B$, which can summarize economically the magnitude and power-law increase in $(q-1)$. We require the straight-line fit  to start at 0.013 at $q=2$ and to have a power-law dependence that can best fit the three points at $q-1 = 2, 3\ {\rm and}\ 4$, as shown by the solid line in Fig.\ 8. The result is
\bq
	A=0.013,    \qquad \qquad  B=4.01.   \label{21}
\eq
These values of $A$ and $B$ represent our current findings for the erraticity indices that characterize quark-hadron PT, as simulated by SCR. The dashed line will be discussed at the end of this section.

	It should be noticed that the scaling region in Fig.\ 7 is for the upper 6 points that correspond to $M>2\times 10^3$ in Fig.\ 6. That region includes the upper region in Fig.\ 1. Thus the erraticity behavior that we are focusing on now is distinct from the intermittency behavior studied in Sec.\ II, not only in the characteristics of scaling, but more obviously in the region where the scaling behavior occurs.
	
	When the data from LHC are analyzed and the values of $\mq$ are determined, the result can be compared with ours from SCR given in Eqs.\ (19) and (19c) or,  in terms of $A$ and $B$ that are more revealing visually, as shown in Fig.\ 9 on $B$ vs $A$. In addition to the point labeled SCR we have added a point for the Ising model, for which
\bq
	{\rm Ising}:  \qquad A=1.2\times 10^{-3}, \qquad  B=2.42   \label{22}
\eq
as given in \cite{ch1}. The drastic difference between Ising and SCR in that plot accentuates the different origins of erraticity of the two systems. For the Ising model there is no time evolution; it has many fluctuating configurations, all of which have clusters of all sizes but no sharp spikes that can give rise to large $\Phi_q(M)$ at large $M$. It therefore has small $\mq$, appearing in the lower-left corner of the plot in Fig.\ 9. On the other hand, SCR can generate particle emissions with large $\Phi_q(M)$ even at large $M$, for which $\mq$ can be large and occupies the upper-right corner of the $A$-$B$ plot, a region that corresponds to high erraticity.

\begin{figure}[tbph]
\vspace*{0.5cm}
\includegraphics[width=.7\textwidth]{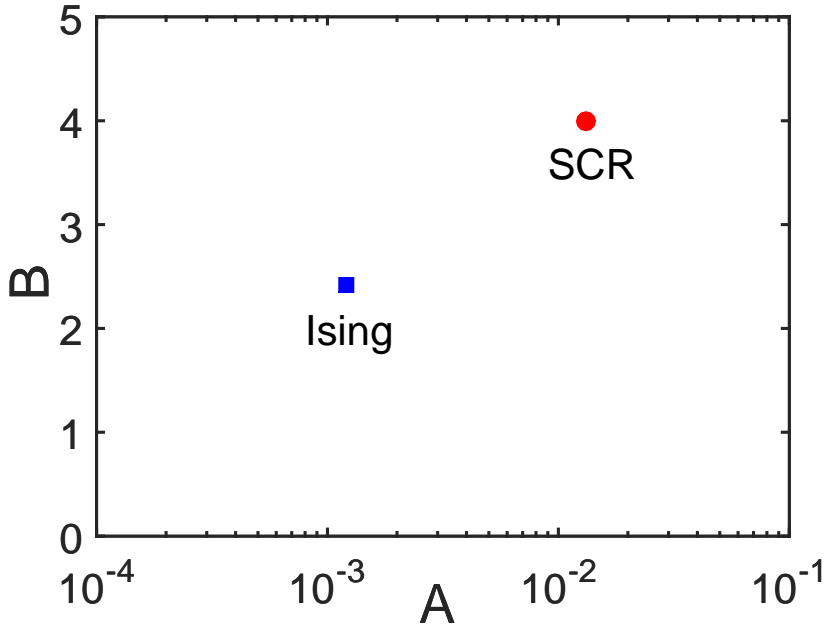}
\caption{(Color online) Values of $A$ and $B$ in Eq.\ (\ref{20}) for SCR (red) and Ising (blue).}
\end{figure}

	Equation (\ref{20}) is a simple power law that makes possible our comparison of SCR with the Ising model shown in Fig.\ 9, but there is no fundamental significance in that particular form. The apparent saturation of $\mq$ at higher $q$ in Fig.\ 8 suggests that a slight modification of Eq.\ (\ref{20}) can represent the values of $\mq$ better. Toward that end we use
\bq
	\mu_q=A' \left({q-1\over q+1}\right)^{B'}   \label{23}
\eq
and show that with the values $A'=51.6$ and $B'=7.51$ the corresponding curve in Fig.\ 8 is the dashed line that fits the points better. Of course, a more elaborate formula with more parameters can always be found to yield even more superior agreement, but that is unnecessary for our purpose here.

Studies of erraticity have been carried out by a number of experiments almost exclusively in nuclear emulsions  \cite{dg, dg1, dg2, aka}. The highest collision energy examined is 200$A$ GeV at CERN SPS, which is still significantly less than what is necessary to avoid averaging over $\phi$ and $\pt$. Comparison of their results on $\mu_q$ with ours would not be too meaningful, since the physics of SCR is not applicable to those experiments. Nevertheless, it may be of interest if those results can be recast in the form of Eq.\ (\ref{20}) and entered as points in Fig.\ 9.

\section{More Analysis on $C_{p,q}$}

In the preceding section we introduced \cpq\ but focused on its properties in the neighborhood of $p=1$. 
Now we apply Eq.\ (\ref{13}) to the high $M$ region in Fig.\ 5  and determine $\psi_q(p)$ for the whole range of $p$ up to 2;  the result is shown in Fig.\ 10 . Evidently, the dependence on $p$ is very nearly linear, as noted earlier above Eq.\ (\ref{14}). Thus the slopes at $p=1$ defined in (\ref{14}) contain properties of $\psi_q(p)$ for all $p<2$. Similar behaviors are found in the lower $M$ regions but will not be exhibited here. Our conclusion on $\mq$ summarized in Figs.\ 8 and 9 is therefore a simple yet substantial representation of erraticity.

\begin{figure}[tbph]
\vspace*{0.5cm}
\includegraphics[width=.7\textwidth]{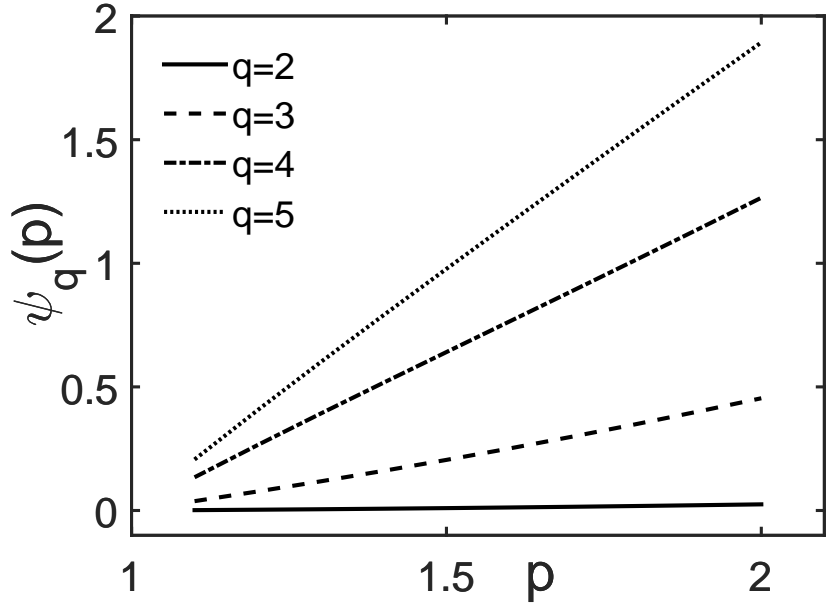}
\caption{Scaling exponents $\psi_q(p)$ vs $p$ for four values of $q$.}
\end{figure}
\begin{figure}[tbph]
\vspace*{0.5cm}
\includegraphics[width=.7\textwidth]{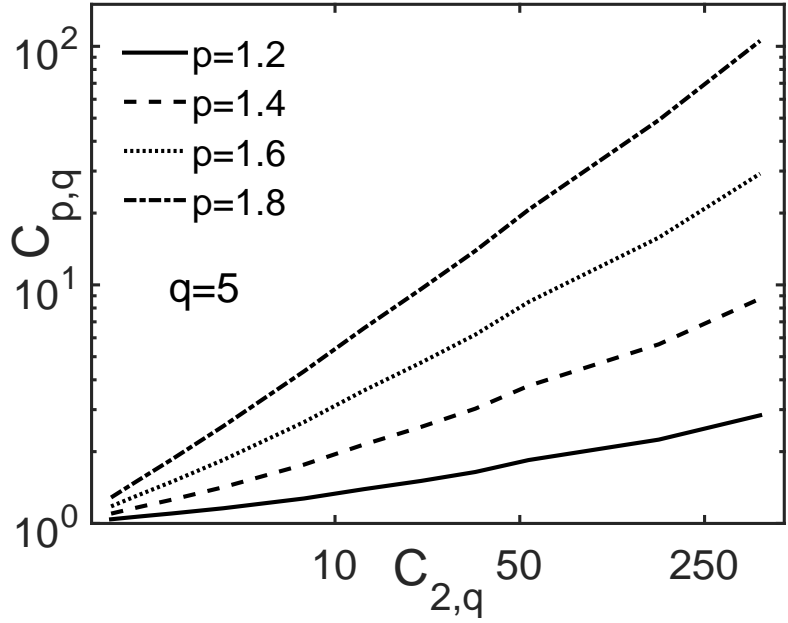}
\caption{Scaling behavior of $C_{p,q}$ vs $C_{2,q}$.}
\end{figure}
\begin{figure}[tbph]
\vspace*{0.5cm}
\includegraphics[width=.7\textwidth]{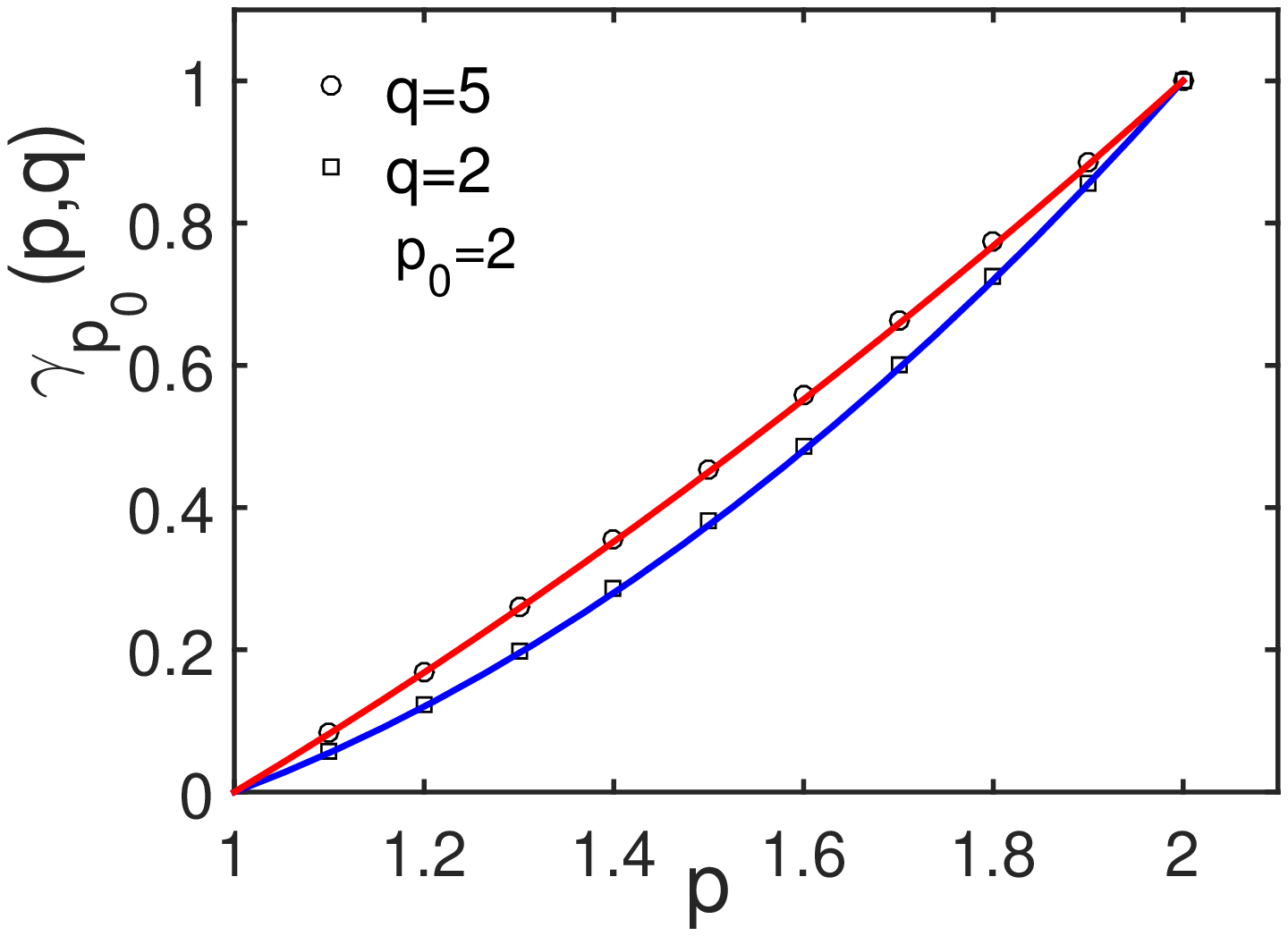}
\caption{(Color online) $\gamma_2(p,q)$ vs $p$ for two values of $q$. The two lines are fits by the same formula Eq.\ (\ref{25}) for $p_0=2$.}
\end{figure}

	Despite the non-uniform behavior of \cpq\ in Fig.\ 5, we can identify simpler features by examining the $ F$-scaling type of properties in $\cp$ vs $C_{p_0,q}$
\bq
\cp(M) \pr C_{p_0,q}(M)^{\gamma_{p_0}(p,q)} .    \label{24}
\eq
This relationship is shown in Fig.\ 11 for $p_0=2$ and $q=5$. Evidently there is a power-law behavior characterized by $\gamma_{p_0}(p,q)$, which is shown in Fig.\ 12 for $q=2, 5$. The lines are fits of the points by the formula
\bq
\gamma_{p_0}(p,q)=(p-1)[1-(p_0-p)/q] ,  \label{25}
\eq
which provides a connection among all $p$ in the interval $1\le p \le 2$. In regions of $M$ where both Eqs.\ (\ref{13}) and (\ref{24}) are valid, we have
\bq
\psi_q(p)=\gamma_{p_0}(p,q) \psi_q(p_0) ,   \label{26}
\eq
which has no explicit $M$ dependence, but depends strongly on the validity of scaling. It is a relationship that can be checked experimentally.

	In Eq.\ (\ref{14}) we have defined $\mq$ to be the derivative at $p=1$. We can extend the definition to any $p$ and write
\bq
\mu_q(p)={d\over dp}\psi_q(p) .   \label{27}
\eq
Using Eq. (\ref{26}) in (\ref{27}), or from seeing that the local tangents of the lines in Fig.\ 12 are all higher than the initial slope at $p=1$, one can conclude that $\mu_q(p) > \mu_q$ for all $p>1$.
The average of $\mq(p)$ in the interval $1\le p \le p_0$ is
\bq
\bar\mu_q = {1\over p_0-1} \int_1^{p_0} dp \mu_q(p) = {\psi_q(p_0)\over p_0-1} .  \label{28}
\eq
Using Fig.\ 10 for $\psi_q(p_0)$  at $p_0=2$ so that $\bar\mu_q=\psi_q(2)$, we obtain the dependence of $\bar\mu_q$ on $q-1$ shown in Fig.\ 13. In adopting the formula
\bq
\bar\mu_q=\bar A (1-{1\over q})^{\bar B} ,   \qquad \bar A=17.7,  \qquad \bar B=9.38 ,   \label{29}
\eq
to describe its behavior, we get the solid line in Fig.\ 13. Note that Eq.\ (\ref{29}) is very nearly, but not exactly, in the form of Eq.\ (\ref{23}). 

\begin{figure}[tbph]
\vspace*{0.5cm}
\includegraphics[width=.7\textwidth]{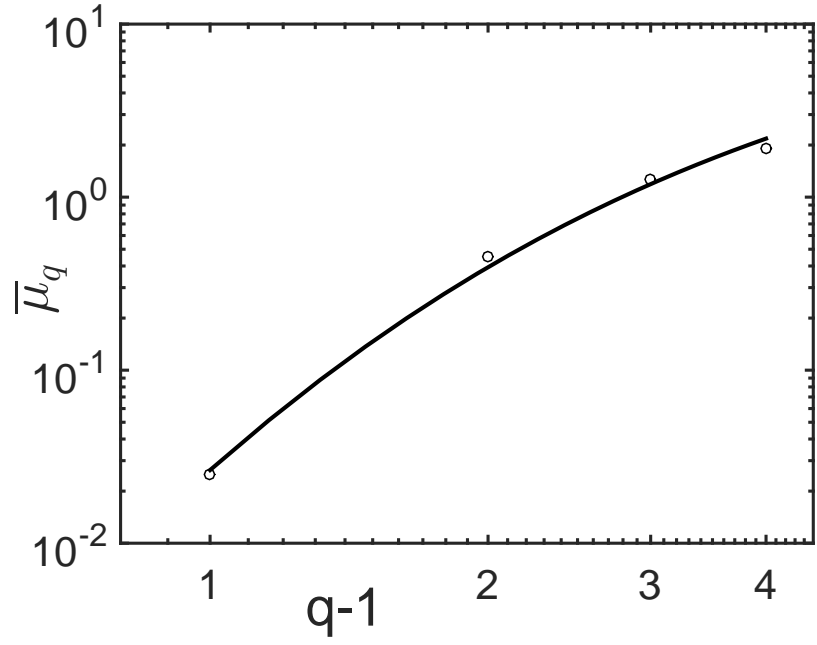}
\caption{The average index $\bar\mu_q$ fitted by Eq.\ (\ref{29}).}
\end{figure}
\begin{figure}[tbph]
\vspace*{0.5cm}
\includegraphics[width=.7\textwidth]{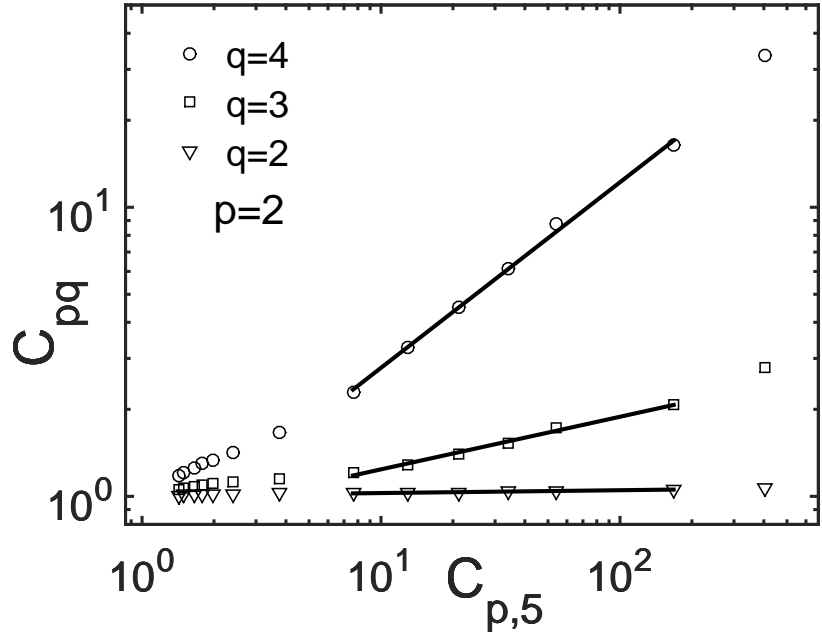}
\caption{Scaling behavior of $C_{p,q}$ vs $C_{p,5}$.}
\end{figure}
\begin{figure}[tbph]
\vspace*{0.5cm}
\includegraphics[width=.7\textwidth]{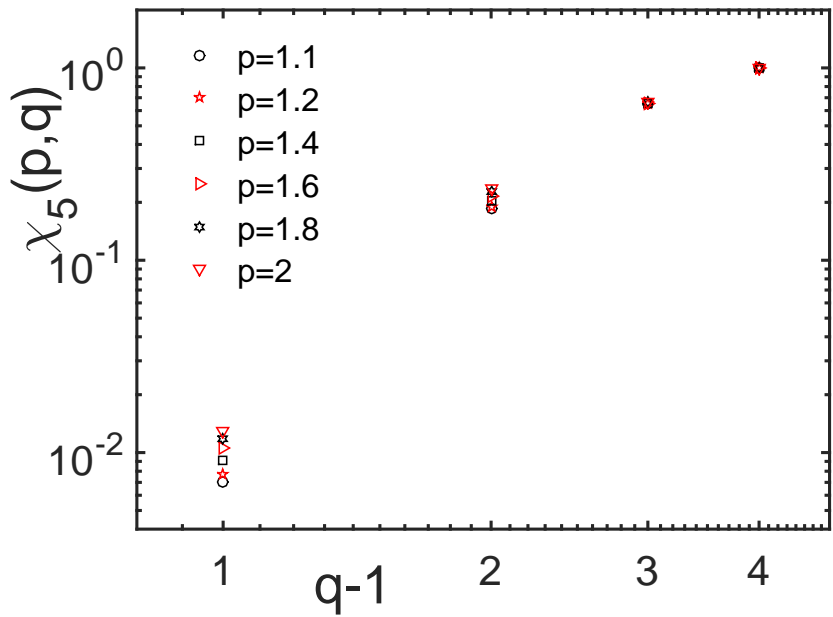}
\caption{(Color online) Scaling exponent $\chi_5(p,q)$ vs $q-1$ for six values of $p$.}
\end{figure}

	Another avenue for the exploration of the scaling behavior of \cpq, beside that of Eq.\ (\ref{24}), is to consider its dependence on $C_{p,q_0}(M)$ with $q_0$ fixed. Figure 14 shows the case for $q_0=5$, obtained from SCR to illustrate the behavior that can be described by
\bq
C_{p,q}(M) \pr C_{p,q_0}(M)^{\chi_{q_0}(p,q)}   \label{30}
\eq
in the scaling region corresponding to large $M$. Although Fig.\ 14 shows only the case $p=2$, the behavior is similar for all $p$ in the whole range of $1<p<2$. Thus it is possible to determine $\chi_{q_0}(p,q)$ that is constrained by its value being 1 at $q=q_0$. The dependency on $q-1$ for various values of $p$ is shown in Fig.\ 15 for $q_0=5$. Evidently, $\chi_5(p,q)$ is not very sensitive to the variation of $p$.
Because the values of $C_{p,q_0}(M)$ for $q_0=2$ vary over limited range, we choose $q_0=5$ for the benefit of maximum effect in the realization of the scaling behavior in Eq.\ (\ref{30}).

	Using the power-law behavior of Eq.\ (\ref{13}) in (\ref{30}), we obtain the relationship that is valid in the scaling region 
\bq
\psi_q(p)=\chi_{q_0}(p,q) \psi_{q_0}(p) .   \label{31}
\eq
Although this equation is distinctly different from Eq.\ (\ref{26}), the similarity of their appearances makes them a companion pair exhibiting different  extrapolations from fixed $p_0$ or fixed $q_0$.

	We can establish contact with our earlier study of properties in the neighborhood of $p=1$ by applying (\ref{31}) to (\ref{14}) and obtain
\bq
\mq=\chi_{q_0}(1,q) \mu_{q_0} ,  \label{32}
\eq
where the condition $\psi_{q_0}(1)=0$ has been used. Referring back to the slope of $\sg$ vs $\Sigma_{q_0}$, which has been denoted by $\omega_{q_0}(q)$ in (\ref{17}), we now have
\bq
\omega_{q_0}(q)=\chi_{q_0}(1,q) .  \label{33}
\eq
To check this relationship we note that it is not possible to obtain $\chi_{q_0}(1,q)$ directly from (\ref{30}) because $\cp=1$ at $p=1$. Figure 15 shows $\chi_5(p,q)$ for various values of $p$ down to $p=1.1$. To compare them to $\omega_5(q)$, we calculate $\omega_5(q)$ by examining $\sg$ vs $\Sigma_5$, which is shown in Fig.\ 16; the slopes of the straight lines are 
\bq
\omega_5(q)=0.0061, 0.136, 0.60, 1.0 \qquad {\rm for} \qquad q=2, 3, 4, 5.  \label{34}
\eq
These values are shown in Fig.\ 17 together with $\chi_5(1.1, q)$, whose values are
\bq
\chi_5(1.1, q)=0.007, 0.184, 0.653, 1.0  \qquad  {\rm for} \qquad q=2, 3, 4, 5.   \label{35}
\eq  
With the expectation that $\chi_5(1, q)$ would be slightly lower than the above, it is remarkable how close (\ref{35}) approaches (\ref{34}) in affirmation of (\ref{33}).

\begin{figure}[tbph]
\vspace*{0.5cm}
\includegraphics[width=.7\textwidth]{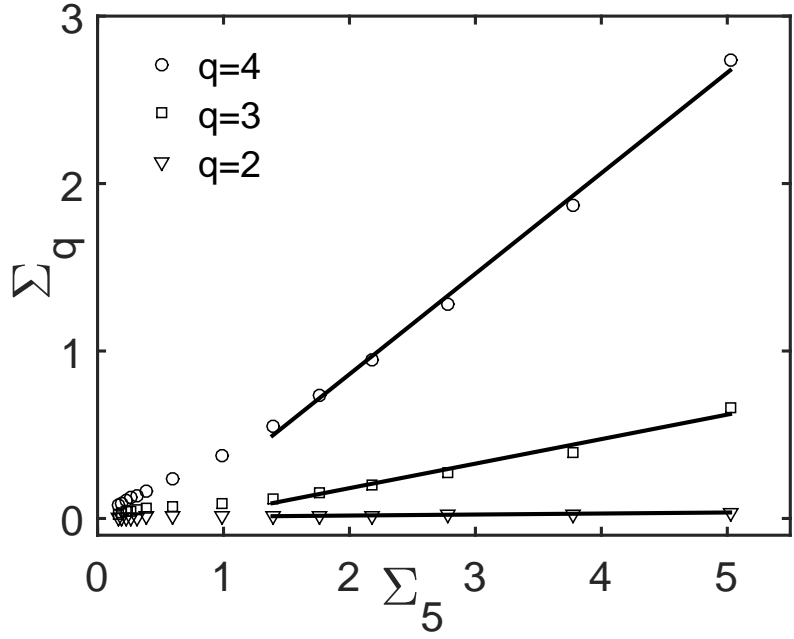}
\caption{Scaling behavior of $\Sigma_q$ vs $\Sigma_5$.}
\end{figure}
\begin{figure}[tbph]
\vspace*{0.5cm}
\includegraphics[width=.7\textwidth]{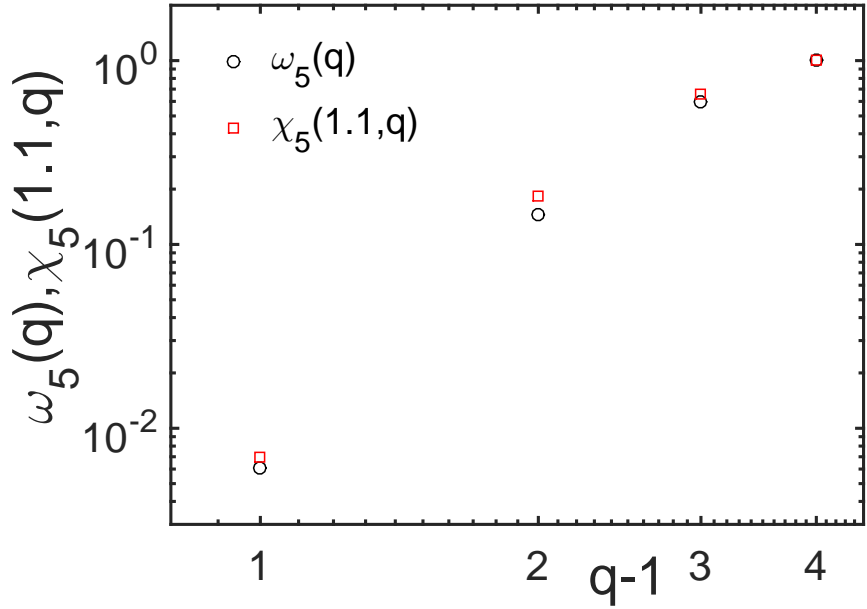}
\caption{(Color online) Comparison of $\omega_5(q)$ (in black) with $\chi_5(1.1,q)$ (in red)}
\end{figure}

	With the combination of (\ref{32}) and (\ref{33}) that gives
\bq
\mu_q=\omega_{q_0}(q) \mu_{q_0} ,   \label{36}
\eq
we can use it for $q_0=5$ to calculate $\mq$ starting with $\mu_5=2.1$ from Eq.\ (\ref{19}). Thus (\ref{34}) implies
\bq
\mu_q=0.0128, 0.307, 1.26   \qquad {\rm for}  \qquad q=2, 3, 4  \label{37}
\eq
in excellent agreement with Eqs.\  (\ref{19c}). It should be recognized that many numerical values of $\omega_2(q)$ and $\omega_5(q)$ have been obtained by power-law fits of points of \cpq\ and $\sg(M)$, generated by SCR, and are not analytically determined. Thus the consistency demonstrated above by different routes of deriving $\mq$ is non-trivial.

\section{Conclusion}

	The properties of quark-hadron phase transition have been studied by use of an event-generator SCR that simulates the dynamics of contraction and randomization of the quark medium in its transition to hadrons. The principal difference between the hadronization process that we study from other schemes, such as fragmentation or recombination, is that our emphasis is on the spatial properties in $(\eta,\phi)$ of the emitted pions instead of their $\pt$ \dis s. Furthermore, the fluctuations of the particle \dis s throughout $(\eta,\phi)$ are the crucial properties that we retain as essential inputs into our analysis for observable signatures of quark-hadron phase transition. In order that the fluctuation behaviors are not overwhelmed by the background, it is necessary to make severe cuts in the admissible $\pt$ range at low $\pt$. That requires the hadron multiplicity to be very high so as to make feasible the fine-grained analysis of the particle \dis\ in  $(\eta,\phi)$. That in turn implies that the collision energy must be very high. For that reason we have entitled this work with reference to LHC, even though no experimental data from LHC have been used.
	
	By use of SCR we have generated multiplicity distributions in as many as $3\times 10^4$ bins so that we can study both the fluctuations from bin to bin in a given event and the event-to-event fluctuations at any fixed bin. We have found that the normalized factorial moments exhibit two scaling regions. In the lower scaling region we concentrated on the intermittency behavior and study the scaling index $\nu$, while in the upper scaling region the focus is on the erraticity index $\mu_q$.
	
	The fact that our result on $\nu$ comes very close to the GL value gives support to the implication that SCR contains the essence of the QCD dynamics responsible for quark-hadron phase transition. When shown in conjunction with the Ising result on $\nu(T)$, we get a broader view of how our SCR result fits into the general scenario of second-order PT where temperature is a control parameter. It suggests a project for the future to include temperature dependence in SCR is such a way that the randomization part is more closely linked to $T$ so as to generate a $T$-dependent $\nu(T)$. The realization of that project will be a challenge due to the fact that $T$ is not an observable. To correlate $T$ to $\pt$ would get us into the more complex domain of analyzing the fluctuation properties in all three kinematical variables  $(\pt,\eta,\phi)$.
	
	The study of the upper scaling region led us to the determination of the erraticity index $\mu_q$. The extensive scaling behavior of the entropy function $\Sigma_q(M)$ vs $\Sigma_2(M)$, shown in Fig.\ 7, convinces us that there is an $M$-independent property that can be extracted. Their slopes (in linear plots) yield $\omega_2(q)$ that facilitates our determination of $\mu_q$. In a simple parametrization of the $q$ dependence of $\mu_q$ we have shown in Fig.\ 9 a comparison of the results from SCR and Ising. It is clear that there are far more fluctuations in SCR (larger $A$ and $B$) than in Ising. Their differences provide a scale to measure the differences of future experimental results compared to SCR, especially those from the analyses of  LHC data.
	
	For the moments-of-moments $C_{p,q}(M)$ we have pushed $p$ into the region between 1 and 2 and found various properties that can be checked by experiments. Since SCR has not been tuned to fit any data, the numerical results of our findings at this stage are not as important as the template that SCR provides in serving as a guide for the directions in which the real data can be analyzed.
	
	Ultimately, the question is what can be learned about quark-hadron phase transition. The first part of our work here makes the contention that the PT is of second order based on the $\nu$ value obtained in SCR being in agreement with Ginzburg-Landau and Ising. But quark-hadron PT may have more properties beyond what GL and Ising contain. If SCR is reliable in generating very erratic fluctuations, then there is a rich territory ahead for real experiments to explore. The study of  such fluctuations at low-$\pt$ has largely been ignored at LHC so far. Experimental investigation of hard jet physics has mainly been following suggestions by theoretical predictions, since QCD is a well-established theory. However, QCD has little to predict at low $\pt$.  Experiments at LHC that focus on the deconfined phase of QCD matter are more concerned about the flow effects, as expected from hydrodynamic models, than about the transition from quarks to hadrons. 	Our event generator SCR is a very crude model that can easily be invalidated in its details by the real data. Thus soft physics is a fertile ground for experiments to lead theory in the development of a realistic description of the physics of confinement in a large system of quarks.
	
 \section*{Acknowledgment}
 
We thank Dr. Edward Grinbaum Sarkisyan for helpful comments. This work was supported in part by the Ministry of Science and Technology of China 973 Grant
2015CB856901 and by National Natural Science Foundation of China under Grant 11435004.

 \end{document}